\documentclass[12pt,a4paper]{article}
\usepackage[truedimen,margin=30mm]{geometry} 

\usepackage{float}
\usepackage{amsmath}
\usepackage{amsthm}
\usepackage{amssymb}
\usepackage{setspace}
\usepackage[authoryear]{natbib}
\usepackage{graphicx}
\usepackage{color}
\usepackage{mathrsfs}
\usepackage{bm}
\usepackage{url}
\usepackage{booktabs}
\usepackage{multicol}
\usepackage{multirow}
\usepackage{titlesec}
\titleformat*{\section}{\large\bfseries}
\titleformat*{\subsection}{\it}
\titleformat*{\subsubsection}{\it}

\newtheorem{cor}{Corollary}
\newtheorem{prp}{Proposition}

\newtheorem{algo}{Algorithm}

\newcommand{\argmin}{\mathop{\rm arg~min}\limits}

\title{{\bf 
Learning rate selection via weighted Fisher divergence
}}
\date{}

\begin{document}

\maketitle
\doublespacing

\vspace{-1.5cm}
\begin{center}
{\large  Takahiro Onizuka
}
\end{center}

\noindent
\begin{center}
    Graduate School of Social Sciences, Chiba University, Japan
\end{center}

\medskip
\medskip
\medskip
\begin{center}
{\bf \large Abstract}
\end{center}

The general Bayesian approach provides a flexible framework by introducing a loss-based likelihood with a learning rate controlling the relative weight of the likelihood and prior. This flexibility creates an important calibration problem under model misspecification, where the conventional Bayesian information identity fails and uncalibrated Gibbs posteriors may not provide reliable uncertainty quantification. We propose a method for selecting the learning rate by minimizing a weighted Fisher divergence between the asymptotic general posterior and a normal distribution with sandwich-type covariance, yielding a closed-form calibrated learning rate. The proposed framework includes Fisher information matching as a special case and, for an important choice of the weight matrix, yields a learning rate no larger than the Fisher information matching rate. This leads to a more dispersed posterior and can reduce the risk of undercoverage. The framework also includes a learning rate based on Kullback--Leibler divergence as a special case. Numerical examples show that the proposed method tends to mitigate undercoverage under model misspecification, while real-data analyses illustrate its practical applicability.

\vspace{-0cm}


\newpage
%

\section{Introduction}

Bayesian inference provides a coherent framework for quantifying uncertainty by combining prior information with evidence from observed data through a likelihood function. In the classical Bayesian paradigm, the likelihood is assumed to represent the data-generating mechanism, at least approximately, and posterior uncertainty is obtained by conditioning on the observed data. This framework is theoretically appealing and practically powerful; however, its validity can be fragile and may lead to misleading posterior uncertainty when the assumed statistical model is misspecified. \cite{bissiri2016general} proposed general Bayesian updating as a coherent loss-based extension of Bayesian inference, in which prior beliefs are updated through a cumulative loss rather than necessarily through a likelihood.
General Bayesian posteriors, also called Gibbs posteriors in many contexts, address this limitation by replacing the negative log-likelihood with a user-specified loss function. Thus, standard Bayes appears as a special case, while the general formulation allows inference to be targeted directly at the parameter or functional defined by the minimizer of an expected loss. This is particularly useful under model misspecification, because the inferential target can be defined operationally through a decision problem rather than through the assumption that a parametric likelihood is true. The general Bayesian approach is widely used because of its flexibility under various risk functions \citep{jiang2008gibbs,bhattacharya2022gibbs,rigon2023generalized} and divergence-based losses \citep[][]{ghosh2016robust,matsubara2022robust,nakagawa2020robust,onizuka2025robust}. In this framework, the learning rate in the loss-based likelihood controls the scale of the loss and plays a crucial role in determining the concentration and uncertainty quantification of the resulting posterior distribution. 

Several approaches have been proposed for selecting or calibrating the learning rate. One line of work uses asymptotic information matching. Under regularity conditions, an empirical risk minimizer often has a sandwich covariance structure involving the expected Hessian of the loss and the covariance of its gradient \citep[see e.g.][]{huber1967behavior}. \cite{lyddon2019general} connected general Bayesian updating with the loss-likelihood bootstrap and proposed selecting the learning rate by matching asymptotic Fisher information. This perspective is attractive because it links the posterior spread to the frequentist variability of the loss-based estimator while preserving a prior-to-posterior belief update. A related perspective arises from the literature on the so-called Safe Bayes approach, proposed by \cite{grunwald2017inconsistency}, which avoids hyper-compression and inconsistency of the general Bayes posterior by selecting a small learning rate. \cite{heide2020safe} also considers the Safe Bayesian approach for generalized linear regression. \cite{holmes2017assigning} proposed selecting the learning rate of power posteriors based on matching the expected information gain between standard Bayesian and general Bayesian updates. \cite{wu2023comparison} provides a direct comparison of several data-driven learning rate selection methods and shows that no single method uniformly resolves severe misspecification, although \cite{syring2019calibrating}'s method tends to perform well for credible-region coverage in their experiments. \cite{syring2019calibrating} constructs calibrated credible sets by selecting a learning rate. This procedure is computationally intensive because it requires repeated MCMC sampling for bootstrap samples. In addition, the calibration method of \cite{syring2019calibrating} depends on a prespecified nominal coverage level. Therefore, the calibration procedure needs to be repeated when credible regions at different coverage levels are of interest. This may result in different calibrated posterior distributions for the same model, depending on the target coverage level.
\cite{agnoletto2025bayesian} provides a general posterior for generalized linear models and consider the selection of the learning rate. Another recent approach to uncertainty quantification under misspecified models was considered by \cite{frazier2023calibrated}. Their approach constructs a Gibbs posterior using a modified loss-based likelihood without selecting a learning rate.

To select a learning rate, we employ a generalization of the Fisher divergence. The Hyv\"arinen score proposed by \cite{hyvarinen2005estimation} is closely related to the Fisher divergence: the expected difference in Hyv\"arinen scores is proportional to the Fisher divergence under suitable regularity conditions. A major advantage of this criterion is that it depends on the model only through derivatives of the log-density, so that normalizing constants cancel out.  This feature makes the Hyv\"arinen score especially useful for unnormalized models and for generalized posterior constructions in which normalizing constants are analytically inconvenient. \cite{matsubara2024generalized} proposed a general posterior based on discrete Fisher divergence and a bootstrap-based calibration procedure that selects the learning rate by minimizing the Hyv\"arinen score. \cite{yang2019variational} compares variational approximations based on the Fisher divergence with those based on the Kullback--Leibler divergence. \cite{huggins2018practical} discuss connections between the Fisher divergence and Wasserstein distance. Weighted versions of the Hyv"arinen score, in which the squared norm is replaced by a weighted norm, have been considered by \cite{barp2019minimum} and \cite{chen2025weighted}.

The proposed approach uses a weighted Fisher divergence under asymptotic normality, which provides a closed-form calibrated learning rate without iterative computation. The resulting framework connects the Fisher information matching learning rate proposed by \cite{lyddon2019general} to weighted scoring rules and provides a unified perspective in which existing calibration criteria can be characterized as special cases. In addition, we propose a procedure based on the Kullback--Leibler divergence, and the resulting learning rate can be obtained as a special case of the weighted Fisher divergence framework with a specific weight. MMoreover, the proposed framework allows the learning rate to be chosen more conservatively from the perspective of uncertainty quantification, thereby reducing the risk of undercoverage relative to the nominal level. In addition, unlike the calibration method of \cite{syring2019calibrating}, the proposed method does not depend on a specific coverage level. Consequently, a single calibrated learning rate can be used to construct a coherent posterior distribution for inference at multiple credible levels of interest.

The rest of the paper is organized as follows. In Sections 2 and 3, we describe the proposed method, its theoretical properties and comparisons with an existing method. Simulation studies evaluating the performance of the proposed method are given in Section 4. In Section 5, we apply the proposed model to real data examples. Section 6 gives concluding remarks.

\section{Methodology}

\subsection{M-estimation}
Let $Z_{1:n}=(Z_1,\dots,Z_n)$ be the observed data, where $Z_1,\ldots,Z_n$ are independent and identically distributed (i.i.d.) observations taking values in $\Omega$ and drawn from an unknown true distribution $P_0$ with density $f_0$. For example, in the standard regression setting with response variable $Y_i$ and covariates $X_i$, each observation is represented by the vector $Z_i=(Y_i,X_i)^\top$. In a parametric framework, we consider a parameterized family of densities $\mathbb{P}_{\Theta}=\{f(\cdot\mid\theta)\mid\theta\in\Theta\subset\mathbb{R}^d\}$, where $\theta$ is the parameter of interest. We define the pseudo-true parameter $\theta_0$ as
\begin{equation}\label{def:theta_0}
    \theta_0=\theta(P_0)=\argmin_{\theta\in\Theta}\int_{\Omega} \ell (\theta,z)dP_0(z),
\end{equation}
where $\ell : \Theta\times \Omega\to[0,\infty)$ is a specified loss function. The model is said to be well specified if $f_0(\cdot)=f(\cdot\mid\theta_0)$ for some $\theta_0\in\Theta$; otherwise, the model is misspecified. Our interest lies in inference on the parameter $\theta$, and we often consider minimizing an empirical risk. Let $\hat{\theta}_n$ be the empirical risk minimizer defined by
$$
\hat{\theta}_n=\argmin_{\theta\in\Theta}\frac{1}{n}\sum_{i=1}^n\ell(\theta,Z_i)=\argmin_{\theta\in\Theta} R_n(\theta),
$$
where $R_n(\cdot)$ denotes the empirical risk. In the frequentist framework, $\hat{\theta}_n$ is commonly used as a point estimator. Under standard regularity conditions, $\hat\theta_n$ is consistent for $\theta_0$ and satisfies asymptotic normality:
\begin{equation}\label{asym-M-est}
  \sqrt{n}(\hat\theta_n-\theta_0)\to N(0,\Sigma_{\mathrm{sand}}),
\end{equation}
in distribution as $n\to\infty$, where the asymptotic covariance matrix is given by
\begin{equation*}
  \Sigma_{\mathrm{sand}}=J(\theta_0)^{-1} I(\theta_0) J(\theta_0)^{-1}=J^{-1} I J^{-1},
\end{equation*}
with
\begin{equation}\label{def:JI}
  J(\theta)=\int_{\Omega} \nabla^2 \ell(\theta,z)dP_0(z),\quad
  I(\theta)=\int_{\Omega} \nabla \ell(\theta,z)\nabla \ell(\theta,z)^\top dP_0(z).
\end{equation}
This asymptotic covariance matrix, called the sandwich covariance matrix, is well known in the robust statistics literature \citep[see, for example, ][]{huber1967behavior}.

\subsection{General Bayesian posterior}

The foundational framework for Bayesian updating with respect to the pseudo-true parameter $\theta_0$ in \eqref{def:theta_0} was established by \cite{bissiri2016general}, who discuss the coherence properties of Bayesian updating under loss functions.
Generalized Bayesian methods focus on inference using the generalized, or Gibbs, posterior of $\theta$. Given a prior distribution $p(\theta)$ and observations $Z_{1:n}$, the generalized posterior distribution is defined as
$$
p_{\mathrm{GB},\omega}(\theta\mid Z_{1:n})\propto p(\theta)\exp\{-\omega n R_n(\theta)\}=p(\theta)\exp\left\{-\omega \sum_{i=1}^n\ell(\theta,Z_i)\right\},
$$
where $\omega>0$ is called a learning rate. In the special case where $\omega=1$ and $\ell(\theta,z)=-\log f(z\mid\theta)$, this formulation reduces to the standard Bayesian framework. Thus, the generalized Bayesian approach extends the classical setup, allowing inference without relying strictly on the assumption that the true data-generating process belongs to the parametric model class. The learning rate $\omega$ controls the scale of the loss function relative to the prior. As $\omega\approx 0$, the exponential influence of the data vanishes, making the posterior identical to the prior. Consequently, $\omega$ plays a key role in balancing the information from the prior and the data.

To determine the learning rate, \cite{lyddon2019general} propose selecting $\omega$ by matching the Fisher information of the asymptotic loss-likelihood bootstrap posterior with that of the generalized posterior. We call the learning rate proposed in \cite{lyddon2019general} the Fisher information matching learning rate, and it is defined by 
\begin{equation}\label{lyddon-omega}
\omega_{\mathrm{FI}}=\frac{\mathrm{tr}\{J(\theta_0)I(\theta_0)^{-1}J(\theta_0)\}}{\mathrm{tr}\{J(\theta_0)\}},
\end{equation}
where $J(\theta_0)$ and $I(\theta_0)$ are defined in \eqref{def:JI}. 
In practice, the method uses empirical estimates of $J(\theta_0)$ and $I(\theta_0)$ by replacing $\theta_0$ with the empirical risk minimizer.
Alternatively, \cite{syring2019calibrating} propose a generalized posterior calibration algorithm based on repeated MCMC sampling to achieve nominal frequentist coverage probabilities. The method aims to construct credible regions that achieve nominal frequentist coverage through repeated MCMC runs over bootstrap samples. \cite{matsubara2024generalized} proposed general Bayesian inference based on discrete Fisher divergence and provided a calibration procedure for the learning rate. This approach focuses on score matching between Bayesian bootstrap--based posterior and the generalized posterior. However, the method may be computationally unstable because it relies on bootstrap-based estimates, and theoretical properties of the method have not yet been fully established.

\subsection{Calibrating the general posterior by weighted score matching}\label{sec:2.3}

As a Bernstein--von Mises--type approximation for generalized posteriors \citep[see e.g.][]{chernozhukov2003mcmc, van2000asymptotic, miller2021asymptotic}, the centered and rescaled generalized posterior satisfies
\begin{equation}
    \sqrt{n}(\theta-\theta_0)\to N\left(0,\,\omega^{-1} J(\theta_0)^{-1}\right)
  \label{eq:bvm_gp}
\end{equation}
in distribution under regularity conditions. We focus on the asymptotic distribution as in \cite{lyddon2019general}. To determine the learning rate, we employ a weighted Fisher divergence. Following \cite{barp2019minimum}, the weighted Fisher divergence between densities $p$ and $q$ is defined as
$$
F_{M}(p\| q)=\int p(x)\|\nabla\log p(x)-\nabla\log q(x)\|_M^2dx,
$$
where $\|\cdot\|_M^2$ is the $M$-weighted norm defined by $\|z\|_M^2=z^\top Mz$ with a positive semidefinite matrix $M$. If $M=I_d$, where $I_d$ denotes the $d$-dimensional identity matrix, then the weighted Fisher divergence reduces to the standard Fisher divergence considered in \cite{hyvarinen2005estimation}. 
The core idea of our proposal is to select $\omega$ such that the asymptotic general posterior \eqref{eq:bvm_gp} minimizes the weighted Fisher divergence relative to the asymptotic frequentist distribution with sandwich covariance. Specifically, we aim to achieve
$$
\omega_{\mathrm{WS}}=\argmin_{\omega} F_{M}(p_{\mathrm{sand}}\|p_{\mathrm{GP},\omega}^{\infty}),
$$
where $p_{\mathrm{sand}}$ and $p_{\mathrm{GP},\omega}^{\infty}$ denote the densities of the multivariate normal distributions corresponding to \eqref{asym-M-est} and \eqref{eq:bvm_gp}, respectively. Note that the weighted Fisher divergence between two multivariate normal distributions is also considered in \cite{chen2025weighted}, where the authors study variational approximation under the $M$-weighted Fisher divergence between $q(\theta\mid z)=N(\theta\mid \nu,\Lambda^{-1})$ and $p(\theta\mid\mu,\Sigma)=N(\theta\mid\mu,\Sigma)$, for which they derive
$$
F_M(p\|q)=\mathrm{tr}(\Sigma^{-1}M)+\mathrm{tr}(\Lambda M\Lambda \Sigma)-2\mathrm{tr}(M\Lambda)+(\mu-\nu)^\top\Lambda M\Lambda (\mu-\nu).
$$
For details, refer to \cite{chen2025weighted}. Using this result, the proposed learning rate based on $M$-weighted Fisher divergence is given in Proposition~\ref{prp:proposed-omega}.

\begin{prp}\label{prp:proposed-omega}
Assume $p_{\mathrm{sand}}(\theta)=N(0,J^{-1}IJ^{-1})$ and $p_{\mathrm{GP},\omega}^{\infty}(\theta)=N(0,\omega^{-1}J^{-1})$. Then, the $M$-weighted Fisher divergence between $p_{\mathrm{sand}}$ and $p_{\mathrm{GP},\omega}^{\infty}$ is minimized by
\begin{equation}\label{proposed-omega}
\omega_{\mathrm{WS}}=\argmin_{\omega} F_{M}(p_{\mathrm{sand}}\|p_{\mathrm{GP},\omega}^{\infty})=\frac{\mathrm{tr}(JM)}{\mathrm{tr}(IM)}.
\end{equation}
\end{prp}

We refer to this selection procedure as \emph{weighted score matching} and to the resulting learning rate as the \emph{weighted score matching learning rate}. The proposed learning rate depends on the metric $M$, and here we consider two choices of the weight matrix $M$. The proposed learning rates for these two choices are given in Corollaries~\ref{cor:I} and \ref{cor:J}. 

\begin{cor}\label{cor:I}
Assume $M=I_d$. The weighted score matching learning rate is given by
$$
\omega_{\mathrm{WS}}=\frac{\mathrm{tr}(J)}{\mathrm{tr}(I)}.
$$
\end{cor}

\begin{cor}\label{cor:J}
Assume $M=J^{-1}$. The weighted score matching learning rate is given by
$$
\omega_{\mathrm{WS}}=\frac{d}{\mathrm{tr}(IJ^{-1})}.
$$
\end{cor}

These corollaries are derived directly from Proposition~\ref{prp:proposed-omega} under $M=I_d$ or $M=J^{-1}$. The $I_d$ weight corresponds to standard score matching and is a natural choice. The $J^{-1}$-weighted score matching criterion has an alternative interpretation in terms of a linear transformation. Specifically, weighting the Fisher divergence by $J^{-1}$ in the original parameterization is equivalent to using the standard Fisher divergence after the transformation by $J^{1/2}$.
\begin{equation}\label{eq:trans-J}
    \omega_{\mathrm{WS}}=\argmin_{\omega} F_{J^{-1}}(p_{\mathrm{sand}}\|p_{\mathrm{GP},\omega}^{\infty})=\argmin_{\omega} F_{I_d}(N(0,K)\|N(0,\omega^{-1}I_d)),
\end{equation}
where $K=J^{1/2}\Sigma_{\mathrm{sand}}J^{1/2}=J^{-1/2}IJ^{-1/2}$.
The two expressions in Corollaries~\ref{cor:I} and \ref{cor:J} are generally different. The difference arises because Fisher divergence is not invariant under linear reparameterization unless the metric is transformed accordingly. Therefore, the metric choice determines the geometry in which score discrepancies are judged. Interestingly, the choice $M=J^{-1}$ also has a connection to the Kullback--Leibler (KL) divergence defined by
$$
\mathrm{KL}(p\| q)=\int p(x)\log\frac{p(x)}{q(x)}dx.
$$
In fact, minimizing the KL divergence between the two asymptotic distributions yields the same learning rate as the $J^{-1}$-weighted score matching criterion in Corollary~\ref{cor:J}.

\begin{prp}\label{prp:KL-omega}
Assume $p_{\mathrm{sand}}(\theta)=N(0,J^{-1}IJ^{-1})$ and $p_{\mathrm{GP},\omega}^{\infty}(\theta)=N(0,\omega^{-1}J^{-1})$. Then, the KL divergence between $p_{\mathrm{sand}}$ and $p_{\mathrm{GP},\omega}^{\infty}$ is minimized by
\begin{equation}\label{KL-omega}
\omega_{\mathrm{KL}}=\argmin_{\omega} \mathrm{KL}(p_{\mathrm{sand}}\|p_{\mathrm{GP},\omega}^{\infty})=\frac{d}{\mathrm{tr}(IJ^{-1})}.
\end{equation}
\end{prp}

Because the KL-based learning rate in Proposition~\ref{prp:KL-omega} coincides with the weighted score matching learning rate for $M=J^{-1}$ in Corollary~\ref{cor:J}, any theoretical property of the weighted score matching learning rate that holds for arbitrary $M$ also applies to the KL-based learning rate. In particular, the KL-based learning rate satisfies Proposition~\ref{prp:well-model} and Corollary~\ref{cor-1d} below.

The weighted score matching approach recovers the standard Bayesian learning rate under correct model specification, as shown in the following proposition. 

\begin{prp}\label{prp:well-model}
If the true density satisfies $f_0(x)=\exp\{-\omega_0\ell(\theta_0,x)\}$ for some $\theta_0\in\Theta$ and $\omega_0>0$, then $\omega_{\mathrm{WS}}=\omega_0$ for any weight $M$.
\end{prp}

This proposition confirms that the proposed rule recovers the learning rate associated with the correctly specified model when the true data-generating process belongs to the assumed parametric family. This desirable property is also satisfied by the Fisher information matching procedure. Note that this property does not depend on the choice of the weight matrix $M$. 

The weighted score matching learning rate in \eqref{proposed-omega} is an oracle quantity because it depends on the unknown parameter $\theta_0$. In practice, we replace the unknown quantities with their empirical counterparts using the empirical risk minimizer $\hat{\theta}_n$ and the empirical estimators of $I$ and $J$ defined in \eqref{def:JI}:
\begin{equation*}
J_n(\theta)=\frac{1}{n}\sum_{i=1}^n\nabla^2\ell(\theta,Z_i),\quad I_n(\theta)=\frac{1}{n}\sum_{i=1}^n \nabla \ell(\theta,Z_i)\nabla\ell(\theta,Z_i)^\top.
\end{equation*}
By the continuous mapping theorem, $\hat{\omega}_{\mathrm{WS}}$ is consistent as $n\to\infty$, provided that $J_n(\hat{\theta}_n)\to J(\theta_0)$ and $I_n(\hat{\theta}_n)\to I(\theta_0)$ in probability.
The practical implementation of the proposed method proceeds as follows:
\begin{algo} 
Given the empirical risk $R_n(\theta)$, proceed as follows.
\begin{enumerate}
  \item Compute the empirical risk minimizer $\hat\theta_n$ and the estimates of $J$ and $I$ using their empirical counterparts $\hat J_n=J_n(\hat\theta_n)$ and $\hat I_n=I_n(\hat\theta_n)$. 
  \item Calculate the weighted score matching learning rate $\hat\omega$.  
  \item Sample from the calibrated generalized posterior:
  $$
    p_{\mathrm{GB},\hat{\omega}}(\theta\mid Z_{1:n}) \propto \exp\{-\hat\omega nR_n(\theta)\}p(\theta).
  $$
\end{enumerate}
\end{algo}

\section{Comparison with existing methods}

\subsection{Analytical comparison with Fisher information matching}\label{sec:3.1}

To compare the proposed method with the Fisher information matching method, we first consider a one-dimensional parameter, i.e. $d=1$. The weighted score matching (WS) and Fisher information matching (FI) learning rates, which are defined in \eqref{proposed-omega} and \eqref{lyddon-omega}, respectively, are denoted by $\omega_{\mathrm{WS}}$ and $\omega_{\mathrm{FI}}$. For a one-dimensional parameter, the weighted score matching learning rate coincides with the Fisher information matching learning rate, as shown in Corollary~\ref{cor-1d}.

\begin{cor}\label{cor-1d}
If the dimension of the parameter $\theta$ is $d=1$,  $\omega_{\mathrm{WS}}=\omega_{\mathrm{FI}}$ for any weight $M$. 
\end{cor}

\cite{wu2023comparison} compare several calibration methods for $d=1$ using a numerical toy example in Section 4 of their paper. Their results show that Fisher information matching can account for model misspecification and achieve approximately nominal coverage in this example. Since $\omega_{\mathrm{WS}}=\omega_{\mathrm{FI}}$ when $d=1$, the weighted score matching procedure necessarily exhibits the same behavior in this setting. Note that, under correct model specification, the weighted score matching and Fisher information matching learning rates coincide for any $d\ge1$ and any weight matrix $M$, as shown in Proposition~\ref{prp:well-model}. For $d\ge 2$, however, the weighted score matching and the Fisher information matching do not coincide in general, as shown in Proposition~\ref{prp:smaller}.

\begin{prp}\label{prp:smaller}
The weighted score matching learning rate with weight $M=I_d$ is no larger than the Fisher information matching learning rate; namely, $\omega_{\mathrm{WS}}\le \omega_{\mathrm{FI}}$.
\end{prp}

\cite{grunwald2017inconsistency} argued that a sufficiently small learning rate can mitigate certain model misspecification biases. Proposition~\ref{prp:smaller} implies that the weighted score matching posterior is asymptotically at least as dispersed as the Fisher information matching posterior. This increased dispersion may reduce undercoverage relative to Fisher information matching.

\begin{cor}
    Assume $M=J^{-1/2}K^{-1/2}JK^{-1/2}J^{-1/2}$ with $K=J^{-1/2}IJ^{-1/2}$. The weighted score learning rate is given by
    $$
    \omega_{\mathrm{WS}}=\frac{\mathrm{tr}\{JI^{-1}J\}}{\mathrm{tr}\{J\}}=\omega_{\mathrm{FI}}.
    $$
\end{cor}

The Fisher information matching learning rate can therefore be represented as a weighted score matching learning rate by choosing $J^{-1/2}K^{-1/2}JK^{-1/2}J^{-1/2}$, where $K$ is the transformed covariance matrix of the sandwich distribution $N(0,K)$ in \eqref{eq:trans-J}. Equivalently, in the transformed parameterization in \eqref{eq:trans-J},
\begin{equation*}
    \omega_{\mathrm{FI}}=\argmin_{\omega} F_{K^{-1/2}JK^{-1/2}}(N(0,K)\|N(0,\omega^{-1}I_d)).
\end{equation*}
Therefore, the $M$-weighted score matching framework includes the Fisher information matching learning rate as a special case, although this metric depends on both $I$ and $J$ and may therefore be less natural than the simple choices $M=I_d$ and $M=J^{-1}$.

\subsection{Analytical example under a normal model with quadratic loss}\label{sec:3.2}

We demonstrate the difference between the weighted score matching and Fisher information matching procedures through a multivariate normal example. Following the setup of \cite{lyddon2019general}, suppose that we observe independent data from a $d$-dimensional multivariate normal distribution, $Y_i\sim N_d(\mu_0,\Sigma_0)$ ($i=1,\dots,n$), and employ a quadratic loss function of the following form:
$$
\ell(\theta,Y_i)=\frac{1}{2}(Y_i-\theta)^\top\Sigma_1^{-1}(Y_i-\theta).
$$
This setup yields $I(\theta_0)=\Sigma_1^{-1}\Sigma_0\Sigma_1^{-1}$ and $J(\theta_0)=\Sigma_1^{-1}$. The Fisher information matching learning rate defined by \eqref{lyddon-omega} is obtained as
$$
\omega_{\mathrm{FI}}=\frac{\mathrm{tr}(\Sigma_0^{-1})}{\mathrm{tr}(\Sigma_1^{-1})}.
$$
Specifically, if we consider the standard quadratic loss where $\Sigma_1=I_d$ and the components of $Y_i$ are independent with $\Sigma_0=\mathrm{diag}(\sigma_1^2,\dots,\sigma_d^2)$, the Fisher information matching learning rate simplifies to $\omega_{\mathrm{FI}}=d^{-1}\sum_{i=1}^d\sigma_i^{-2}$. Under our framework, the weighted score matching learning rates given in Corollaries~\ref{cor:I} and \ref{cor:J} both reduce to $\omega_{\mathrm{WS}}=d/\sum_{i=1}^d\sigma_i^{2}$. If all variances are equal, then $\omega_{\mathrm{WS}}=\omega_{\mathrm{FI}}$. Otherwise, the Fisher information matching learning rate is larger than the weighted score matching learning rate, even in this simple setup.

\subsection{Illustration under a two-dimensional model}\label{sec:3.3}

Consider a bivariate normal model $Y_i\sim N_2(0, \mathrm{diag}(\sigma_1^2,\sigma_2^2))$ $(i=1,\dots,100)$ with the standard quadratic loss $2^{-1}(Y_i-\theta)^\top(Y_i-\theta)$ for estimating the population mean. The Fisher information matching learning rate is given by $\hat{\omega}_{\mathrm{FI}}=2^{-1}(\hat{\sigma}_1^{-2}+\hat{\sigma}_2^{-2})$ and the weighted score matching learning rate is given by $\hat{\omega}_{\mathrm{WS}}=2/(\hat{\sigma}_1^{2}+\hat{\sigma}_2^{2})$, where $\hat{\sigma}_i^2$ ($i=1,2$) denote the empirical variances. We consider two distinct data-generating settings: (A) $\sigma_1^2=\sigma_2^2=1$; and (B) $\sigma_1^2=1,\sigma_2^2=5$. We assign the priors $\theta\sim N_2(0,\tau^2I_2)$ and $\tau^2\sim\mathrm{IG}(0.1,0.1)$, where $\mathrm{IG}(a,b)$ denotes the inverse gamma distribution with shape parameter $a$ and scale parameter $b$. 

\begin{figure}
    \centering
    \includegraphics[width=1.0\linewidth]{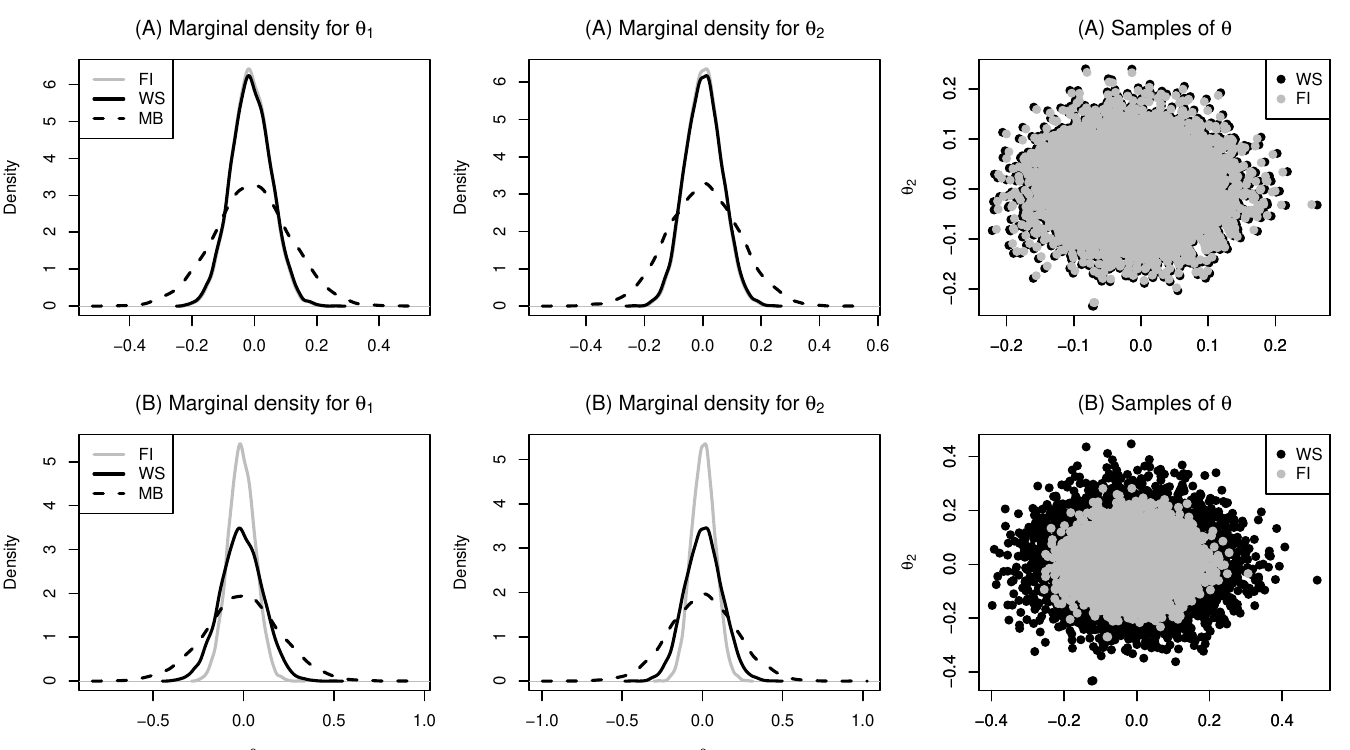}
\caption{Plots of the posterior distributions from Section~\ref{sec:3.3}. The left and center panels show the posterior densities obtained via WS (black line), FI (gray line), and MB (dashed line). In the right panels, the black and gray points represent posterior samples obtained using WS and FI, respectively.}
    \label{fig:2D-example}
\end{figure}

The resulting posterior distributions are illustrated in Figure~\ref{fig:2D-example}. As shown in the top panels for the isotropic case (A), the posteriors from both methods are nearly identical, consistent with the theoretical result in Section~\ref{sec:3.1}. However, in the presence of variance heterogeneity (B), the two methods yield noticeably different posteriors. For comparison, we also include the misspecified Bayesian (MB) model based on a Gaussian likelihood with precision parameter $\omega$ and gamma prior for $\omega$ in the left and center panels.

\section{Numerical example}

We focus on learning-rate selection for the general posterior and compare the proposed weighted score matching procedures defined in Corollaries~\ref{cor:I} and \ref{cor:J} with existing methods in terms of coverage probability. We refer to the procedures in Corollaries~\ref{cor:I} and \ref{cor:J} as WS and KL, respectively, because the latter coincides with the KL matching procedure established in Proposition~\ref{prp:KL-omega}.
Although \cite{syring2019calibrating} directly deals with the calibration of credible regions and is designed to achieve nominal frequentist coverage of credible regions, it can incur a large computational cost. Therefore, in this section, we compare the proposed method only with the Fisher information matching procedure proposed by \cite{lyddon2019general}. The bootstrap calibration proposed by \cite{syring2019calibrating} is employed only in the real-data application in Section~\ref{sec:5}.

\subsection{Least Squares Regression with Quadratic Loss}\label{sec:4.1}

Let $y_1,\dots,y_n\in\mathbb{R}$ and $x_{1},\dots,x_n\in\mathbb{R}^d$ be the response values and covariate vectors, respectively. The regression mean for observation $i$ is modeled as $x_i^\top\theta$, where $\theta$ is an unknown regression parameter. To estimate the mean function, we use the quadratic loss
$$
R_n(\theta)=\frac{1}{n}\sum_{i=1}^n\ell(\theta,z_i)=\frac{1}{n}\sum_{i=1}^n(y_i-x_i^\top\theta)^2.
$$
The data-generating models $y_i=x_i^\top\theta+\epsilon_i$ are as follows:
\begin{itemize}
\item (Case 1) $\epsilon_i\sim N(0,1)$.
\item (Case 2) $\epsilon_i\sim N(0,\sigma_i^2)$, where the standard deviation $\sigma_i$ is defined by
$$
\sigma_i=
    0.25 \cdot 1_{\{x_{i1}<\hat{x}_{0.05}\}}+
    0.5 \cdot 1_{\{\hat{x}_{0.05}\le x_{i1}\le\hat{x}_{0.95}\}} +
    1_{\{x_{i1}>\hat{x}_{0.95}\}},
$$
where $\hat{x}_{0.05}$ and $\hat{x}_{0.95}$ denote the sample 5th and 95th percentiles of $x_{11},\dots,x_{n1}$. 
\item (Case 3) $\epsilon_i\sim t_3$, where $t_\nu$ is the $t$-distribution with degrees of freedom $\nu$.
\end{itemize}
The $d$-dimensional covariates $x_i$ are generated independently from a multivariate normal distribution with mean zero, unit marginal variances, and an AR(1)-type correlation structure $E[x_{ij}x_{ik}]= 0.2^{|j-k|}$. We set the true coefficient vector to be $\theta=(1,1,2,-1)$ and $n=100, 400,$ and $1000$. Cases 2 and 3 follow the settings considered in \cite{wu2023comparison} and represent severe misspecification.
The misspecified Bayesian model (denoted by MB), using a Gaussian likelihood and a gamma prior $\mathrm{Ga}(0.1,0.1)$ for the precision parameter, is also included for comparison. For all methods, we assume $\theta\sim N_d(0,\tau^2I_d)$, $\tau^2\sim \mathrm{IG}(0.1,0.1)$.

\subsection{Robust regression with Huber loss}\label{sec:4.2}

To address the presence of outliers, the Huber loss is often used:
$$
R_n(\theta)=\frac{1}{n}\sum_{i=1}^n\rho_c(y_i-x_i^\top\theta), \quad \rho_c(u)=\begin{cases}
    \frac{1}{2}u^2 & |u|\le c\\
    c|u|-\frac{1}{2}c^2 & |u|>c
\end{cases},
$$
where $y_1,\dots,y_n\in\mathbb{R}$ and $x_{1},\dots,x_n\in\mathbb{R}^d$ are response variables and covariates, respectively. Here, $c>0$ is a tuning parameter for robustness and a commonly used choice is $c=1.345$. We implement posterior sampling using the \texttt{rstan} package in R. We use the same data-generating models as those in Section~\ref{sec:4.1}.

\subsection{Results}

For each method, we generate 7000 posterior samples and discarded the first 2000 samples as burn-in. To assess performance, the coverage probability (CP) of the 95\% highest posterior density (HPD) credible region is estimated over 1000 Monte Carlo replications. Tables~\ref{tab:sim-LS-CP} and \ref{tab:sim-Huber-CP} summarize the average estimated learning rates and the empirical coverage probabilities over 1000 replications.
All methods perform reasonably well in Case 1, where model misspecification is relatively mild. The two proposed methods maintain coverage above 90\% in all settings and generally achieve coverage close to the nominal level of 95\%, and the estimated values $\hat{\omega}$ of the two proposed method are also similar. The FI method, however, has lower CP values, particularly in Case 2. This is consistent with Proposition~\ref{prp:smaller}, because FI tends to select a larger learning rate than the WS method, resulting in a more concentrated posterior. under severe misspecification scenarios. In these simulations, the estimated learning rates generally decrease with increasing $n$. Note that the mean squared errors of the posterior mean estimators are essentially identical across methods.

\begin{table}[htbp]
    \caption{Average estimated learning rates and empirical coverage probabilities of the 95\% highest posterior density credible regions over 1000 replications for quadratic loss.}
    \begin{center}
    \begin{tabular}{cc|cccc||cccc}
    \toprule
    \multirow{2}{*}{Case} & \multirow{2}{*}{$n$} &  \multicolumn{4}{c||}{$\hat{\omega}$} & \multicolumn{4}{c}{CP$\times100$ (\%)}\\
         &  & WS & KL & FI & MB & WS & KL & FI & MB \\
         \midrule
         & 100 & 0.55 & 0.55 & 0.59 & 0.51 & 92.70 & 92.80 & 90.70 & 94.20 \\ 
        1 & 400 & 0.52 & 0.52 & 0.53 & 0.50 & 93.30 & 93.40 & 92.90 & 94.10 \\ 
         & 1000 & 0.51 & 0.51 & 0.51 & 0.50 & 96.00 & 95.90 & 95.90 & 95.70 \\ 
         \midrule
         & 100 & 8.41 & 8.49 & 22.12 & 11.30 & 91.90 & 91.60 & 45.30 & 81.10 \\ 
        2 & 400 & 5.57 & 5.65 & 11.09 & 9.04 & 93.60 & 92.70 & 71.60 & 77.40 \\ 
         & 1000 & 5.18 & 5.27 & 8.57 & 8.73 & 92.20 & 92.00 & 78.60 & 77.10 \\ 
                  \midrule
         & 100 & 0.23 & 0.23 & 0.28 & 0.21 & 93.20 & 93.50 & 85.10 & 95.90 \\ 
        3 & 400 & 0.19 & 0.19 & 0.22 & 0.19 & 95.20 & 95.30 & 91.80 & 95.20 \\  
         & 1000 & 0.18 & 0.18 & 0.20 & 0.18 & 94.20 & 94.00 & 91.70 & 93.60 \\ 
         \bottomrule
    \end{tabular}
    \end{center}
    \label{tab:sim-LS-CP}
\end{table}

\begin{table}[htbp]
\caption{Average estimated learning rates and empirical coverage probabilities of the 95\% highest posterior density credible regions over 1000 replications for Huber loss.}
\begin{center}
    \begin{tabular}{cc|ccc||ccc}
    \toprule
    \multirow{2}{*}{Case} & \multirow{2}{*}{$n$} &  \multicolumn{3}{c||}{$\hat{\omega}$} & \multicolumn{3}{c}{CP$\times100$ (\%)}\\
         &  & WS & KL & FI  & WS & KL & FI  \\
         \midrule
         & 100 & 1.26 & 1.23 & 1.37 & 92.00 & 92.40 & 90.00 \\ 
        1 & 400 & 1.19 & 1.18 & 1.22 & 94.10 & 94.10 & 93.80 \\ 
         & 1000 & 1.17 & 1.17 & 1.18 & 96.00 & 96.00 & 96.00 \\ 
         \midrule
         & 100 & 18.00 & 17.86 & 52.19 & 93.10 & 93.40 & 47.10 \\ 
        2 & 400 & 13.89 & 13.92 & 25.38 & 93.70 & 93.60 & 76.20 \\ 
         & 1000 & 13.46 & 13.55 & 20.84 & 91.80 & 91.80 & 80.70 \\ 
                  \midrule
         & 100 & 0.96 & 0.92 & 1.05 & 93.20 & 93.90 & 90.80 \\  
        3 & 400 & 0.90 & 0.89 & 0.92 & 94.40 & 94.20 & 93.80 \\  
         & 1000 & 0.89 & 0.89 & 0.90 & 94.90 & 95.00 & 94.90 \\
         \bottomrule
    \end{tabular}
\end{center}
    \label{tab:sim-Huber-CP}
\end{table}

\section{Application}\label{sec:5}

\subsection{Housing price data}\label{sec:House}

We consider the housing price data set {\tt hprice1} from the {\tt wooldridge} package in R, which contains information on 88 houses sold in the Boston area in 1990. The data include the selling price of each house together with several housing characteristics. In this analysis, we use the selling price (price, measured in thousands of dollars) as the response variable and the lot size (lotsize, in square feet), floor area (sqrft, in square feet), and number of bedrooms (bdrms) as explanatory variables. We standardize these variables before the analysis. An important feature of this data set is the presence of heteroskedasticity. In particular, the variability of housing prices tends to increase with housing characteristics such as lot size and floor area, suggesting that the conditional variance of the error term may depend on the explanatory variables. This behavior is plausible for housing data because expensive or large houses can exhibit substantial price variation due to unobserved factors, such as location, construction quality, neighborhood characteristics, and other amenities that are not explicitly included in the regression model. Indeed, this housing price regression has been used as a standard example of heteroskedasticity in econometric analysis, and tests based on squared OLS residuals provide evidence against the assumption of constant error variance \citep{wooldridge2009introductory}. Therefore, this data set provides a useful empirical example for investigating regression methods that allow the conditional error variance to vary with the covariates, rather than imposing the conventional homoskedastic assumption. \cite{berenguer2021heteroscedasticity} also use the same linear regression specification as an empirical example of heteroskedasticity. 
We employ the quadratic loss and compare the two proposed methods with the Fisher information (FI) matching method, the bootstrap-based calibration (BC) method proposed by \cite{syring2019calibrating} and the misspecified model (MB) using Gaussian likelihood. For the BC method, we set the target coverage level to 95\% because this calibration procedure requires a prespecified nominal coverage level, and the details of our implementation of the BC calibration procedure are summarized in the Appendix. For all methods, we generate 7000 posterior samples, discard the first 2000 as burn-in, and use the remaining 5000 samples for posterior inference.

\begin{figure}
    \centering
    \includegraphics[width=1.0\linewidth]{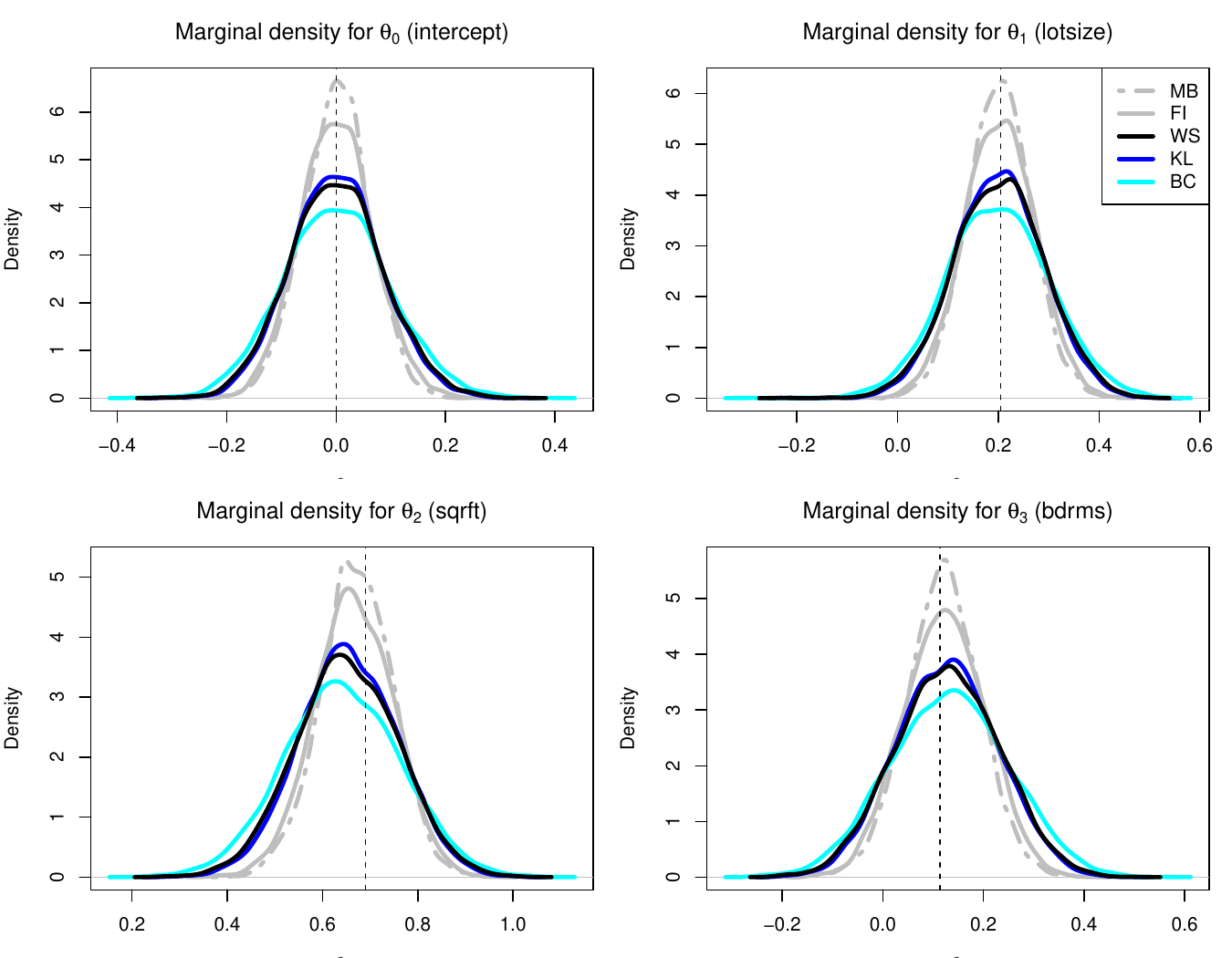}
\caption{Posterior densities obtained by the five methods: FI (gray line), WS (black line), KL (blue line), BC (cyan line) and MB (dashed gray line). The dotted vertical lines represent the frequentist point estimates $\hat{\theta}_n$.}
    \label{fig:boston}
\end{figure}

The estimated learning rates and computation times (seconds) are reported in Table~\ref{tab:application}, and the estimated posterior densities are shown in Figure~\ref{fig:boston}. Note that the estimated $\hat\omega$ in the MB method is given by the posterior mean of the precision parameter, and the computation time includes the procedure for selecting $\hat{\omega}$ and Gibbs sampling using the selected $\hat{\omega}$. The BC method gives the smallest value. In addition, the WS method yields the second-smallest estimated learning rate although the estimated learning rates and posterior distributions obtained by WS and KL are similar. On the other hand, both the MB method and the FI method provide tighter posteriors than the other methods. The proposed approach yields more dispersed posterior distributions than the FI method, which is consistent with its tendency to reduce undercoverage observed in the simulation study. In terms of computational cost, although the computation times of WS, KL, and FI are short, the BC method requires a much longer time to select the learning rate due to the repeated MCMC runs required for calibration. Note that the computation time of the BC method depends on the initial value and step size.

\subsection{Biochemists data}\label{sec:Bio}

In this section, we apply the proposed methods to the biochemists data from \cite{long1990origins}, which contain 915 observations and six variables: the number of publications by Ph.D. biochemists, gender (fem), marital status (mar), the number of children (kid5), the prestige of the Ph.D. program (phd) and the number of articles published by the mentor during the previous three years (ment). The {\tt bioChemists} dataset is available in the {\tt pscl} package for R. We set the number of publications as the response variable and the other variables as covariates. Since the response variable is count-valued, we use the negative log-likelihood of the Poisson distribution as the loss function and assign a normal prior to the regression coefficients. Note that the standard Poisson regression model fails to account for the excess zeros observed in this dataset and is therefore likely to be misspecified \citep[e.g.,][]{long2001predicted}. MCMC sampling is implemented using the \texttt{rstan} package. We compare the two proposed methods with the FI and BC methods and with the misspecified Bayesian (MB) model using $\omega=1$. We use the same computational settings as in Section~\ref{sec:House}. For all methods, we generate 7000 posterior samples, discard the first 2000 as burn-in, and use the remaining 5000 samples for posterior inference.

\begin{figure}
    \centering
    \includegraphics[width=1.0\linewidth]{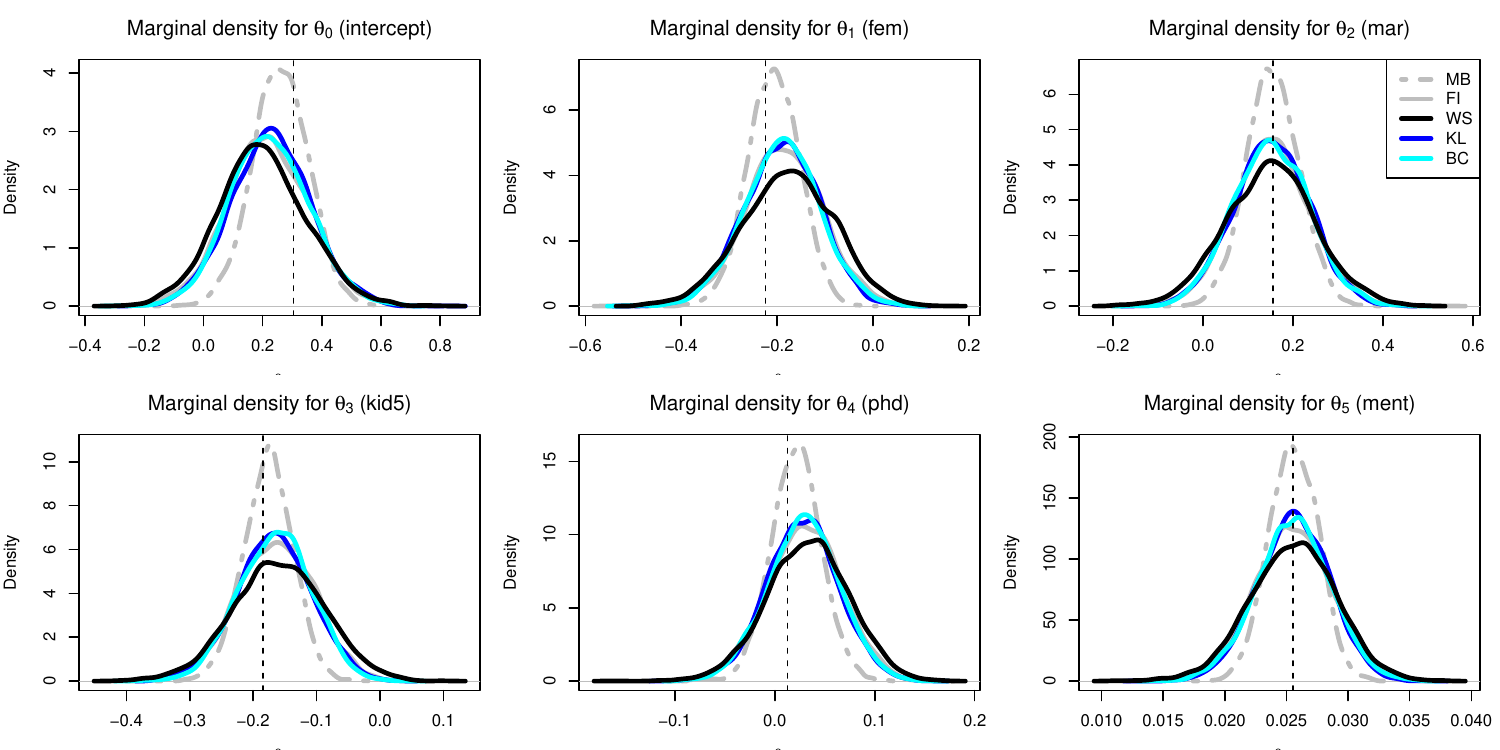}
\caption{Posterior densities obtained by the five methods; FI (gray line), WS (black line), KL (blue line), BC (cyan line) and MB (dashed gray line). The dotted vertical lines represent the frequentist point estimates.}
    \label{fig:application}
\end{figure}

The estimated posterior densities are shown in Figure~\ref{fig:application}. The estimated learning rate and the computation times are summarized in Table~\ref{tab:application}. The estimated learning rates obtained by WS and KL differ in this application. In particular, the KL learning rate is larger than the FI learning rate, whereas the WS learning rate is smaller than the FI learning rate. This illustrates that the ordering result in Proposition~\ref{prp:smaller} applies to WS but not necessarily to KL. Although the estimated learning rates obtained by KL and BC are similar, there is no substantial difference between the resulting posterior distributions. The proposed WS method may give wider posterior distributions and may provide a more conservative learning-rate calibration because it yields the smallest learning rate among WS, KL, and FI in this application. As in Section~\ref{sec:House}, the BC method requires substantially more computation time than the other methods.

\begin{table}[htbp]
    \centering
    \caption{Estimated learning rates and computation times (in seconds).}
    \begin{center}
    \begin{tabular}{c|cccccc}
    \toprule
        Section &  & WS & KL & FI & BC & MB\\
        \midrule
        \multirow{2}{*}{\ref{sec:House}} & Estimated $\hat{\omega}$ & 0.711 & 0.770 & 1.200 & 0.545 & 1.471 \\ 
         & Computation time & 0.186 & 0.157 & 0.152 & 131.669 & 0.182 \\ 
         \midrule
        \multirow{2}{*}{\ref{sec:Bio}} & Estimated $\hat{\omega}$ & 0.315 & 0.456 & 0.408 & 0.445 & 1\\ 
         & Computation time  & 1.76 & 1.69 & 1.66 & 1549.32 & 1.92\\ 
         \bottomrule
    \end{tabular}
    \end{center}
    \label{tab:application}
\end{table}

\section{Concluding remarks}

The proposed learning rate minimizes the weighted Fisher divergence between the asymptotic generalized posterior distribution and the asymptotic sampling distribution with sandwich covariance. The weighted score matching learning rate depends on a weight matrix, and we consider two natural choices motivated by transformations of the asymptotic distribution and a connection to the Kullback--Leibler divergence. The proposed approach can yield smaller learning rates than the Fisher information matching method of \cite{lyddon2019general} and provides uncertainty quantification with coverage probabilities generally close to the nominal level, thereby reducing the risk of undercoverage. We also show that both Fisher information matching and KL matching can be characterized as special cases of the proposed framework. Compared with the calibration method of \cite{syring2019calibrating}, the proposed method does not rely on repeated calculation or prespecified coverage levels and therefore enables computationally efficient calibration based on a single coherent posterior distribution.

There are several directions for future work. First, extending the proposed calibration approach to variational Bayes is also of interest \citep[e.g.][]{onizuka2024fast}. Second, prior information may play an important role in small samples, motivating the development of higher-order asymptotic approximations or non-asymptotic calibration methods \citep[e.g.][]{matsubara2024generalized}. Third, although the proposed method relies on the weighted Fisher divergence, other discrepancy measures or calibration criteria should also be considered.

\section*{Acknowledgments}

The author is supported in part by JSPS KAKENHI Grant Numbers 25K23108 and 26K00322 from the Japan Society for the Promotion of Science.

\appendix

\section{Proof of theoretical property}

\subsection{Proof of Proposition 1}

We obtain
$$
F_M(N(\theta\mid\mu,\Sigma)\|N(\theta\mid \nu,\Lambda^{-1}))=\mathrm{tr}(\Sigma^{-1}M)+\mathrm{tr}(\Lambda M\Lambda \Sigma)-2\mathrm{tr}(M\Lambda)+(\mu-\nu)^\top\Lambda M\Lambda (\mu-\nu)
$$
by a simple calculation \citep[see e.g.][]{chen2025weighted}. Under $p_{\mathrm{sand}}(\theta)=N(0,\Sigma_{\mathrm{sand}})$ and $p_{\mathrm{GP},\omega}^{\infty}(\theta)=N(0,\omega^{-1}J^{-1})$, 
\begin{align*}
    F_{M}(p_{\mathrm{sand}}\|p_{\mathrm{GP},\omega}^{\infty})&=\mathrm{tr}(\Sigma_{\mathrm{sand}}^{-1}M)+\mathrm{tr}(\omega J M\omega J \Sigma_{\mathrm{sand}})-2\mathrm{tr}(M\omega J)\\
    &\propto\omega^2\mathrm{tr}(J MJ \Sigma_{\mathrm{sand}})-2\omega \mathrm{tr}(MJ).
\end{align*}
This is a quadratic function of $\omega$, and the minimizing value of $\omega$ is given by $\mathrm{tr}\{MJ\}/\mathrm{tr}\{JMJ\Sigma_{\mathrm{sand}}\}$ with $\Sigma_{\mathrm{sand}}=J^{-1}IJ^{-1}$.

\subsection{Proof of Proposition 2}

Let $p_{N(\mu,\Sigma)}(\theta) $ and $p_{N(\nu,\Lambda^{-1})}(\theta)$ be the densities of $N(\mu,\Sigma) $ and $N(\nu,\Lambda^{-1})$, respectively. The Kullback Leibler divergence between two Gaussian distributions $N(\mu,\Sigma) $ and $N(\nu,\Lambda^{-1})$ is given by
\begin{align*}
\mathrm{KL}(N(\theta\mid\mu,\Sigma)\|N(\theta\mid \nu,\Lambda^{-1}))&= \int p_{N(\mu,\Sigma)}(\theta) \log\frac{p_{N(\mu,\Sigma)}(\theta)}{p_{N(\nu,\Lambda^{-1})}(\theta)} d\theta \\
&=\frac{1}{2}\left[\log\frac{|\Lambda^{-1}|}{|\Sigma|}+\mathrm{tr}(\Sigma\Lambda)+(\mu-\nu)^\top\Lambda (\mu-\nu)-d\right]
\end{align*}
Under $p_{\mathrm{sand}}(\theta)=N(0,\Sigma_{\mathrm{sand}})$ and $p_{\mathrm{GP},\omega}^{\infty}(\theta)=N(0,\omega^{-1}J^{-1})$, 
\begin{align*}
    f(\omega)=\mathrm{KL}(p_{\mathrm{sand}}\|p_{\mathrm{GP},\omega}^{\infty})&\propto -d\log \omega + \omega\mathrm{tr}\{\Sigma_{\mathrm{sand}}J\}
\end{align*}
Then, $f'(\omega)=-d/\omega + \mathrm{tr}\{\Sigma_{\mathrm{sand}}J\}$, $f''(\omega)=d/\omega^2>0$ and, by solving $f'(\omega)=0$, 
$$
\omega=\argmin_{\omega} \mathrm{KL}(p_{\mathrm{sand}}\|p_{\mathrm{GP},\omega}^{\infty})=\frac{d}{\mathrm{tr}\{\Sigma_{\mathrm{sand}}J\}}=\frac{d}{\mathrm{tr}\{IJ^{-1}\}}. 
$$

\subsection{Proof of Proposition 3}

If the true density satisfies $f_0(x)=\exp\{-\omega_0\ell(\theta_0,x)\}$ for some $\theta_0\in\Theta$ and $\omega_0>0$, then it is well known that $J(\theta_0)=\omega_0 I(\theta_0)$ holds \citep[see e.g.][]{lyddon2019general}. Therefore, $M$-weighted score matching learning rate is given by
$$
\omega_{\mathrm{WS}}=\frac{\mathrm{tr}\{ J(\theta_0)M\}}{\mathrm{tr}\{I(\theta_0)M\}}=\frac{\mathrm{tr}\{\omega_0I(\theta_0)M\}}{\mathrm{tr}\{I(\theta_0)M\}}=\omega_0.
$$

\subsection{Proof of Corollary 3}

For $d=1$, $I=I(\theta_0)$ and $J=J(\theta_0)$ are also scalar. Therefore, the Fisher information matching rate defined by \eqref{lyddon-omega} is given by $\omega_{\mathrm{FI}}=\mathrm{tr}\{JI^{-1}J\}/\mathrm{tr}\{J\}=I^{-1}J$. In contrast, the weighted score matching learning rate $\omega_{\mathrm{WS}}$ for any weight $M$ is given by 
\begin{equation*}
    \omega_{\mathrm{WS}} =\mathrm{tr}\{JM\}/\mathrm{tr}\{IM\}=I^{-1}J
\end{equation*}

\subsection{Proof of Proposition 4}

We consider $I_d$-weighted score matching learning rate. For the Frobenius inner product $\langle A,B\rangle=\mathrm{tr}(A^\top B)$, 
\begin{align*}
    &\langle I^{1/2},I^{-1/2}J\rangle=\mathrm{tr}(I^{1/2}I^{-1/2}J)=\mathrm{tr}(J),\\
    &\langle I^{1/2},I^{1/2}\rangle=\mathrm{tr}(I),\quad \langle I^{-1/2}J,I^{-1/2}J\rangle=\mathrm{tr}(JI^{-1}J).  
\end{align*}
Using the Cauchy--Schwarz inequality $\langle A, B\rangle^2\le \langle A,A\rangle\langle B, B\rangle$, we obtain $\{\mathrm{tr}(J)\}^2\le \mathrm{tr}(I)\mathrm{tr}(JI^{-1}J)$ and then
$$
\omega_{\mathrm{WS}}=\frac{\mathrm{tr}(J)}{\mathrm{tr}(I)}\le \frac{\mathrm{tr}(JI^{-1}J)}{\mathrm{tr}(J)}=\omega_{\mathrm{FI}}
$$
since $\mathrm{tr}(J)>0$, $\mathrm{tr}(I)>0$. The equality in the Cauchy--Schwarz inequality holds if and only if
\(I^{1/2}\) and \(I^{-1/2}J\) are linearly dependent. Hence, there exists
a constant \(c>0\) such that $I^{-1/2}J = c I^{1/2}$,
which is equivalent to $J = cI$.

\section{Details of \cite{syring2019calibrating} in the Application}

\cite{syring2019calibrating} aims to assess the empirical coverage probability through bootstrap samples. The details of the algorithm are as follows.

\begin{algo}
Set the initial value $\omega_{(0)}$ and $t=0$, and resample $B$ bootstrap samples $Z_{1:n}^{(1)},\dots,Z_{1:n}^{(B)}$ based on the original data $Z_{1:n}$.
\begin{enumerate}
\item For $b=1,\dots,B$, construct credible regions $C_{\omega_{(t)},\alpha}(Z_{1:n}^{(b)})$, which satisfy $P_{\mathrm{GP}}(\theta\in C_{\omega_{(t)},\alpha}(Z_{1:n}^{(b)})\mid Z_{1:n}^{(b)})=1-\alpha$. 
\item Evaluate the empirical coverage probability
$$
\hat{c}_{\alpha}(\omega_{(t)})=\frac{1}{B}\sum_{b=1}^B 1\{\hat{\theta}_n\in  C_{\omega_{(t)},\alpha}(Z_{1:n}^{(b)})\}
$$
\item If $|\hat{c}_{\alpha}(\omega_{(t)})-(1-\alpha)|<\epsilon$, then Stop the repitition and select $\hat{\omega}=\omega_{(t)}$; otherwise update
$$
\omega_{(t+1)}=\omega_{(t)}+\kappa_t\{\hat{c}_{\alpha}(\omega_{(t)})-(1-\alpha)\}
$$
and repeat the steps.
\end{enumerate}
\end{algo}

\cite{syring2019calibrating} recommends $\kappa_t=ct^{-0.51}$ for $c>0$, $B=200$ and $\epsilon=1/B$. The initial value $\omega_{(0)}$ and $c>0$ are critical for fast convergence in Step 3. Since it is difficult to search the range of $\hat\omega$, a procedure to set the initial value should be considered.  Here, we set $c=2$ and $\omega_{(0)}=\hat{\omega}_{\mathrm{FI}}$.

\vspace{1cm}
\bibliographystyle{chicago}
\bibliography{refs}

\end{document}